\documentclass[prl,twocolumn,showpacs,amsmath,amssymb]{revtex4-1}

\usepackage{graphicx}% Include figure files
\usepackage{amssymb,amsmath}
\usepackage{dcolumn}% Align table columns on decimal point
\usepackage{bm}% bold math
\usepackage{color}

\begin{document}

\title{Stable magnetic monopoles in spinor polariton condensates}

\author{D. D. Solnyshkov}
\affiliation{LASMEA, Nanostructure and Nanophotonics Group, Clermont Universit\'{e}}
\affiliation{Universit\'{e} Blaise Pascal, CNRS, 63177 Aubi\`{e}re Cedex France}

\author{H. Flayac}
\affiliation{LASMEA, Nanostructure and Nanophotonics Group, Clermont Universit\'{e}}
\affiliation{Universit\'{e} Blaise Pascal, CNRS, 63177 Aubi\`{e}re Cedex France}

\author{G. Malpuech}
\affiliation{LASMEA, Nanostructure and Nanophotonics Group, Clermont Universit\'{e}}
\affiliation{Universit\'{e} Blaise Pascal, CNRS, 63177 Aubi\`{e}re Cedex France}

\date{\today}

\begin{abstract}
We consider half-integer topological defects in a spinor Bose-Einstein condensate of exciton-polaritons. We show that they accelerate under the effect of an in-plane effective magnetic field, like regular electric charges in the electric field. The trajectories of these analogues of magnetic monopoles are derived analytically and checked by numerical simulations. We show that these "magnetic charges" propagate at large in-plane velocities and remain stable up to a critical magnetic field, thanks to the unique spin-anisotropy of polariton interactions.
\end{abstract}

\maketitle

In the famous work of 1931 \cite{Dirac}, which started the quest for magnetic monopoles, P. Dirac considered nontrivial "nodal lines". These objects are characterized by a vanishing wavefunction on these lines and a nonintegrable phase around them. He found that the end of a nodal line is a magnetic pole with a quantized magnetic charge. Twenty years later, Onsager and Feynman \cite{Onsager,Feynman55} described vortex lines in a 3D Bose-Einstein condensate (BEC) which show strong similarities with Dirac's nodal lines. Contrary to the nodal lines, the vortex lines terminate at the boundaries of the BEC. Another important difference between the nodal lines and the vortex lines is that the former are defined for a true single-particle wavefunction, whereas the latter concern the order parameter of the condensate described with a mean-field Gross-Pitaevskii equation \cite{PitaevskiiBook}. Vortices are now much better known to scientists, because they have been nucleated and thoroughly analyzed experimentally \cite{VortexReview}, contrary to magnetic monopoles. 

A scalar condensate does not provide a vector field, which could be arranged in the required pattern, and therefore, an analogue of a Dirac nodal line can only be implemented using topological excitations in a more complex media. The similarity between the spin texture of certain vortices in spinor condensates and the field of a monopole has been rediscovered (after the initial idea of P. Dirac) about 10 years ago \cite{VortexReview,VolovikPNAS}, and proposals have been made in order to organize the required structures in atomic condensates \cite{WFmonopoles,Antiferro,Atomic2009}. Moreover, the textures of similar type have been obtained in different media, such as the spin ice, and it was confirmed that they indeed behave like charges, making possible what was called "magnetricity" in analogy with "electricity" \cite{Magnetricity2009}.

In this Letter, we show the striking analogy between magnetic monopoles and half-integer topological defects in spinor polariton condensates. Polaritons are the quasiparticles formed of photons and excitons strongly coupled in microcavities \cite{Microcavities}. They have recently become a model system for studying condensation \cite{Kasprzak2006} and various associated effects such as superfluidity \cite{AmoSuperfl}, formation \cite{DeveaudScience} and evolution \cite{RecentVortexEvolution} of vortices, oblique solitons \cite{Pigeon,ScienceAmoSoliton}, and others. Their popularity is backed up by their fascinating properties, mixing those of light (small effective mass) and matter (self-interactions and thermalization with phonons). Their spin structure is especially interesting: polaritons are bosons with only two allowed spin projections on the growth axis: $\pm1$. The most important feature of polaritons for the present work is the strong spin-anisotropy of their interactions\cite{ReviewSpin}. Indeed, the exciton-exciton interaction constant in the triplet configuration $\alpha_{1}$ is repulsive and approximately ten times larger than in the singlet configuration $\alpha_{2}$ \cite{Renucci2005}. Consequently, the interaction energy of the polariton condensate is minimized when the latter is linearly polarized\cite{Shelykh2006}: $n_{+}=n_{-}=n/2$ ($n$ is the total density). The spin-anisotropy of interactions makes the condensate stable against small in-plane effective magnetic fields which are acting on the polariton pseudospin \cite{ReviewSpin}, as we will see below.

The two spin components make the existence of half-vortices\cite{RuboHV} and half-solitons\cite{FlayacHalfSoliton} in polariton condensates quite natural. Indeed, the first experimental observation of a half-vortex has been carried out for a polariton condensate\cite{DeveaudScience}. We demonstrate that a constant effective magnetic field (whose origin will be described below) accelerates such half-integer defects, which therefore behave as magnetic charges capable of propagating at high velocities. We demonstrate that these defects remain stable if the field does not exceed a critical value. The 1D half-soliton case is studied first, followed by a more complicated configuration of a half-vortex in a 2D condensate. 

\emph{1D system} Let us consider a two component condensate described by the spinor Gross-Pitaevskii equations (GPE):
\begin{equation}
i\hbar \frac{{\partial {\psi _\sigma }}}{{\partial t}} =  - \frac{{{\hbar ^2}}}{{2 m^*}}{\Delta\psi _\sigma } + \alpha_{1} {\left| {{\psi _\sigma }} \right|^2}{\psi _\sigma }-\frac{\Omega}{2}\psi_{-\sigma},
\end{equation}
where $\sigma=\pm$ and $m^*$ is an effective mass associated with a parabolic dispersion approximation for polaritons, valid at relatively small wavevectors. The term with $\Omega$ describes the linear polarization splitting or in-plane effective magnetic field, mixing the circular components. The spin anisotropy of the polariton-polariton interaction is fully taken into account by using the conditions $\alpha_{1}\gg \alpha_{2}$ and neglecting $\alpha_{2}$ hereafter. For the following discussion, it is convenient to use the pseudospin representation : $\mathbf{S}=(S_x,S_y,S_z)$ with ${S_{x}}=\Re \left( {\psi _{+}\psi _{-}^{\ast }}\right)$, ${S_{y}}=\Im \left( {\psi _{+}^{\ast }\psi _{-}}\right)$, ${S_{z}}=\left( {n_{+}-n_{-}}\right)/2$. Indeed, the pseudospin evolution can be mathematically described by the interaction of this pseudospin with an effective magnetic field, similarly to that of an ordinary magnetic momentum subject to a real magnetic field \cite{ReviewSpin}.
The in-plane effective magnetic field in microcavities can describe for example the energy splitting between the two orthogonal linear polarizations linked with crystallographic axes and it can be controlled by an applied electric field \cite{APLMalpuech}. There is usually an additional LT splitting in 1D wire cavities \cite{Dasbach2005}.

\begin{figure}[t]
\includegraphics[width=0.99\linewidth]{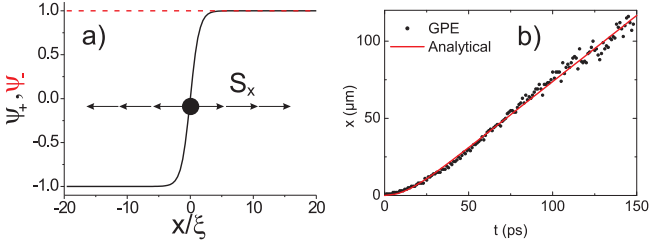}
\caption {(Color online) a) Wavefunction components for a dark half-soliton together with the corresponding in-plane pseudospin pattern; b) Trajectory of a half-soliton accelerated in a constant in-plane effective magnetic field calculated analytically (red solid line) and numerically (black points).}
\label{fig1}
\end{figure}

For a homogeneous solution $\psi_\pm=\sqrt{n/2}$, $\psi_{\mp}=\sqrt{n/2}e^{i\phi_0}$, the relative phase $\phi_0$ between the two components gives the direction of the linear polarization, or the in-plane component of the pseudospin $\mathbf{S}_{\parallel }=(S_x,S_y)$. It is well known that the GPE admits soliton solutions, the latter can lie in only one of the two components of a spinor condensate giving a half-soliton (HS): $\psi_{+}(x)=\sqrt{n/2}\tanh{(x/\xi\sqrt{2})}$ where $\xi=\hbar/\sqrt{\alpha_{1} n m^*}$ is the healing length of the condensate, while the other component can remain constant $\psi_{-}=\sqrt{n/2}e^{i\phi_{0}}$. The in-plane pseudospin in the particular case where $\phi_0=0$ is given by $S_{x}(x)=n\tanh{(x/\xi\sqrt{2})}$, $S_y(x)=0$. Far from the soliton's core of the width $\xi$, the pseudospin has therefore exactly the pattern of the field created by a single magnetic charge in 1D, shown in Fig.\ref{fig1}(a).

Let us analyze in the limit $L\gg \xi$, the effect of a magnetic field on the HS in order to verify that it behaves indeed as a magnetic charge. 
To find the force acting on the HS we have to find the gradient of the magnetic energy of the system as a function of the displacement of the soliton's core. The magnetic energy of a condensate containing a HS in an external in-plane magnetic field is $\int (\mathbf{\Omega/2}\cdot\mathbf{S}) dx$, with $\mathbf{S}(x-x_{0})=n {\rm sign}(x-x_{0})\mathbf{e}_x$ in the limit we consider, giving $E(x)=-\Omega_{x} n x$ and $F=\Omega_{x}n$. For a grey soliton propagating with a speed $v$, the phase shift induced by the soliton is $\Delta S=2 {\rm arc cos}(v/c)<\pi$ and the pseudospin projection $S_x$ is reduced, which can be expressed as a renormalization of the magnetic charge. In the limit $L\gg \xi$, the correction to the charge is found analytically as $q=q_0(1-v^2/c^2)$. The total correction for the mass of the soliton and its charge gives the equation of motion $a=q_{0} \Omega_{x} n/m_{0}(1-v^2/c^2)^{3/2}$, the same as in relativistic physics, integrating which one obtains $v(t)=c \tanh (q_{0} \Omega_{x} n t/c)$ (zero initial velocity). The resulting trajectory is shown in Fig.\ref{fig1}(b), perfectly fitting the results of numerical simulations (see below).

Under the effect of the effective magnetic field the HS can indeed propagate similarly to a real charge, its stability against the pseudospin rotation being maintained by the interaction energy. However, if the in-plane magnetic field exceeds a certain value, the HS can change sign: convert into the other polarization component. The estimation of the critical field $\Omega_{c}$ can be obtained analyzing the dynamical equation for the pseudospin in the center of the HS: $\partial_t\mathbf{S} = \boldsymbol{\Omega}  \times \mathbf{S}$, with $\boldsymbol{\Omega}=\Omega_{x}/\hbar\mathbf{e}_{x}+\alpha_{1} S_z\mathbf{e}_{z}$. The analytical solution of this equation is expressed in terms of elliptical functions demonstrating regular oscillations, and the asymptotic behavior of the $S_z$ pseudospin component at critical field $\Omega_c=\alpha_{1} n_{+}/2$ is exponential. Below $\Omega_c$ the pseudospin keeps the initial sign of its projection over $z$: the HS is stable. In atomic condensates the spin anisotropy is usually almost absent ($\alpha_1\simeq\alpha_2$) and thus the HS would be destroyed. The same result applies to the 2D case for half-vortices as we will see. This dynamical nonlinear effect is different from the spin-Meissner effect obtained for the ground state of polaritonic systems\cite{KavokinMeissner}.

\begin{figure}[h]
\includegraphics[width=0.99\linewidth]{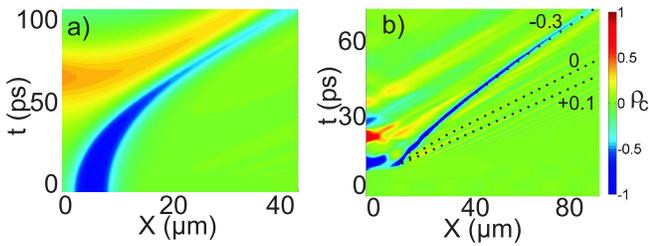}
\caption {(Color online). Circular polarization degree as a function of coordinate and time calculated using the spinor GPE, including the effect of a constant in-plane effective magnetic field. The HS trajectory is visible as the deep blue minimum. a) initial conditions, no lifetime; b) pumping and lifetime, the HS is created by a pulsed potential. Black dashed lines are the guides for the eyes showing trajectories for the splitting values indicated on the figure (in meV).}
\label{fig2}
\end{figure}

To check the analytical predictions, we have performed two types of numerical simulations using the spinor GPE. First, we have imposed a HS as an initial condition and simulated its propagation under the effect of a magnetic field. The resulting circular polarization degree is shown in Fig.\ref{fig2}(a), the blue region corresponding to the soliton. The time dependent position of the soliton has been extracted from this simulation and successfully compared with analytical predictions in Fig.\ref{fig1}(b).

Second, we have performed a more realistic simulation of a wire-shaped cavity with finite lifetime under quasi-resonant, homogeneous, and linearly polarized pumping by an external laser of amplitude $P_\sigma$ and frequency $\omega$:
\begin{equation}
i\hbar \frac{{\partial {\psi _\sigma }}}{{\partial t}} =  - \frac{{{\hbar ^2}}}{{2m^{*}}}\Delta {\psi _\sigma } + \alpha_{1} {\left| {{\psi _\sigma }} \right|^2}{\psi _\sigma } - \frac{{i\hbar }}{{2\tau }}{\psi _\sigma } - \frac{\Omega}{2}\psi_{-\sigma} + {P_\sigma }{e^{ - i\omega t}}
\end{equation}

In this case the HS is created after the steady state regime is obtained, by a short potential pulse acting on a single polarization component $\sigma^+$. Applying this potential on the wire's edge allows creating a single HS propagating at speed $v$ and accelerating as its depth decreases. Thus, the finite lifetime acts as an effective force \cite{PitaevskiiBook}, accelerating a relativistic particle ($m \simeq {m_0}/\sqrt {1 - {v^2}/{c^2}} $) up to the limiting speed, proportional to the speed of sound $c=\sqrt{\alpha_{1} n/2m}$. The effect of the magnetic field can be seen as an extra force $F(t)=q(v(t))\Omega_{x}n$, which can decrease or increase the acceleration [Fig.\ref{fig2}(b)]. We see that even in the driven-dissipative case the magnetic monopoles are sufficiently stable to propagate under the effect of the magnetic field on long distances.

An important difference between the Dirac's monopoles and the half-integer defects in spinor condensates lies in the interactions between them. While a real magnetic charge creates a real magnetic field which certainly influences other magnetic charges, half-integer defects possess only a (pseudo)spin field, which has a weak influence on other half-integer defects. They behave indeed as magnetic charges in an external magnetic field, but see each other rather via interaction and kinetic energy.

\emph{2D system}
In the 2D case the object possessing the spin texture close to that of a magnetic monopole is the elementary half-vortex (HV) carrying winding number $\pm 1$ in the component where it appears. However, there is an important difference between the field of an electric charge, which in 2D decreases as $1/r$, and the pseudospin "field" of the HV, which is approximatively constant at large distances. In spite of that, it is still possible to find the force acting on the HV from the magnetic field as the gradient of the energy of the system as the function of the displacement of the monopole.

\begin{figure}[h]
\includegraphics[width=0.99\linewidth]{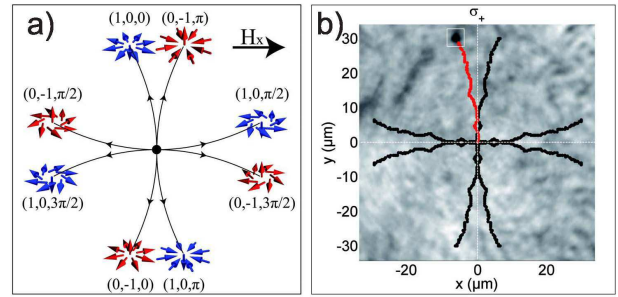}
\caption {(Color online). a) Pseudospin textures for HVs as a function of topological charges and relative phase $\phi_{0}$ (labeled as ($l_+,l_-,\phi_0$)) and the resulting propagation directions in a constant magnetic field pointing along the $x$ direction. b) The $\sigma^{+}$ density at the end of the simulation with the Gross-Pitaevskii equation showing the trajectory (red line) of a particular HV (white square) together with the other trajectories (black lines) corresponding to the cases described on a).}
\label{fig3}
\end{figure}

A crucial peculiarity arises from the relative phase $\phi_0$ of the two wavefunction components: $\psi_{+}=\sqrt{n_{+}}e^{i(l_+\phi+\phi_0)}$, $\psi_{-}=\sqrt{n_{-}}e^{i(l_-\phi)}$ ($\phi$ is the polar angle) which gives an extra degree of freedom for the pseudospin orientation. Several pseudospin textures are shown in Fig.\ref{fig3}(a) together with the resulting sketched trajectory of the HV in the constant in-plane magnetic field $H_x$ (black arrow). These trajectories have been calculated by solving the 2D spinor GP equations with infinite lifetime. We have considered a cylindrical trap of radius 50 $\mu$m with impenetrable boundaries and we have found the associated HV "ground state" at fixed chemical potential $\mu=1$ meV with a given set ($l_+,l_-,\phi_0$). Then, for each of the 8 HV configurations shown on the Fig. 3(a) we have applied a constant magnetic field along the x-direction. The resulting calculated trajectories are shown in black on the Fig. 3(b). Experimentally, a half-vortex can be created in a cylindrical trap (mesa) with a Gauss-Laguerre laser beam.

We see that the propagation direction is determined both by the topological charges ($l_+$,$l_-$) of the HV and by the continuous variable $\phi_0$.  We note that the $(l_+,l_-,\phi_0)$ and the $(-l_+,-l_-,\phi_0)$ HVs are symmetric with respect to the magnetic field and follow the same trajectory. Thus, the proper description for the force acting on the HV from the magnetic field is given by a charge tensor: $F_{i}=q_{ij}H_{j}$, where $q_{ij}$ are proportional to components of the rotation matrix for angle $(\pi-\phi_{0})$ (for the conventions used above).

In order to understand better the forces acting on the HV, let us consider the (1,0,0) HV, whose trajectory is shown in red and whose propagation is demonstrated in the supplementary material \cite{movie}. Two effects induced by magnetic field, are observed in the simulations.

\begin{figure}[h]
\includegraphics[width=0.99\linewidth]{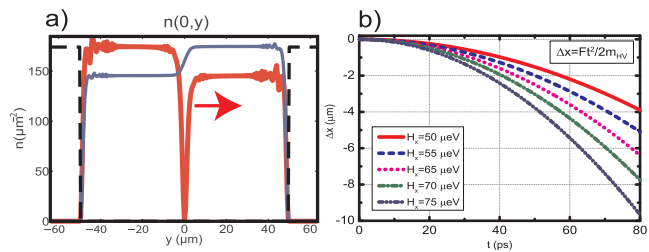}
\caption {(Color online). a) The density distributions of the two components: $\sigma^{+}$ (red) and $\sigma^{-}$ (blue), creating the gradient of the interaction energy accelerating the HV in the $y$ direction. b) The fitted $x$-coordinate of the HV as a function of time for several values of the magnetic field, demonstrating constant acceleration of the half-vortex, increasing with the field.}
\label{fig4}
\end{figure}

First, because of the pseudospin rotation around the magnetic field far from the vortex core, a density gradient appears at short times along the field's axis ($x$ in our case) as shown on the [Fig.\ref{fig4}(a)]. The density determines the vortex energy, whose gradient creates an additional force acting on the HV. Fortunately, this force is strong and well defined only in the initial moments after the application of the field, because later on the different eigenfunctions of the vessel oscillate with different frequencies and the resulting "storm in a teacup" gives zero net force for the vortex, which therefore propagates with a constant speed acquired in the initial moments.

Second, the constant effect of the magnetic field on the pseudospin texture creates a constant force accelerating the vortex in the direction given by eq. (3). The trajectory is parabolic, with the acceleration proportional to the strength of the magnetic field, as shown on the [Fig.\ref{fig4}(b)]. It is precisely this effect, and not the previous one, which corresponds to the expected behavior of a magnetic monopole. 
The HV is remarkably stable against density fluctuations and we are able to track its trajectory for 80 ps. Relativistic effects are not observed in this case, because the speed $v_{x}$ remains relatively small.

 We note that the effect of a static polarization splitting on a HV has been analyzed in Ref.\cite{RuboPolSplit} and a polarization string has been predicted to be attached to the HV in the lowest energy state. This is indeed what we observe in the early moments of the simulation, when the magnetic field is turned on \cite{movie}, but the string tends to unbind from the HV because our initial condition corresponds to the absence of the field and this initial condition is not the eigenfunction of the complete Hamiltonian, otherwise it would not evolve with time.

The HV can become unstable against magnetic field, with the stability criterion given in our case by the same estimate as in 1D. If the magnetic field is stronger, the vortex disappears. Without spin-anisotropy of interactions in the homogeneous magnetic field we consider, the pseudospin along the $y$ axis would make a turn around the magnetic field $H_{x}$ and become completely $S_{z}$ on one side and $-S_{z}$ on the other side, leading to the destruction of the HV. Roughly speaking, instead of having a single point with zero density of $\sigma^{+}$, we would have a semi-infinite string. This is the signature of the fact that in the case of a spinor condensate, the velocity circulation is not conserved topologically, because one cannot define unique superfluid velocity\cite{Leggett}. Therefore, the HVs can be destroyed (unwound) by applying a certain magnetic field, as it has been demonstrated experimentally\cite{Leggett2}. In the case of polaritons, the particular spin-dependence of interactions prevents this from happening for sufficiently low magnetic fields.

Finally, we would like to note that the relative phase between the condensate components can be modified by applying a weak external magnetic field in the $z$ direction, leading to a Zeeman splitting for the $\sigma^{+}$ and $\sigma^{-}$ components. The relative phase would increase linearly with time and therefore the direction of the in-plane magnetic force would change as well.

\emph{Conclusions} Half-integer topological defects in spinor polariton condensates behave as magnetic charges with respect to an in-plane effective magnetic field. Their stability is provided by the spin-anisotropy of polariton-polariton interactions. We have shown that the dynamics of half-solitons in 1D is well described by relativistic equations.

We would like to thank M. Glazov for fruitful discussions. We acknowledge the support of FP7 ITN "Spin-Optronics" (237252), ANR "Quandyde", and IRSES "POLAPHEN" (246912) projects.


\begin{thebibliography}{99}

\bibitem{Dirac} P. A. M. Dirac, \emph{Proc. Roy. Soc. A} \textbf{133}, 60 (1931).

\bibitem{Onsager} L. Onsager, \emph{Nuovo Cimento} \textbf{6}, 249 (1949).

\bibitem{Feynman55} R. P. Feynman, \emph{Progr. Low Temp. Phys.} \textbf{I}, 17 (North-Holland, Amsterdam, 1955).

\bibitem{PitaevskiiBook}  L. P. Pitaevskii, and S. Stringari, \emph{Bose-Einstein Condensation}, Cambridge University Press, Cambridge, (2003).

\bibitem{VortexReview} K. Kasamatsu, M. Tsubota, M. Ueda, \emph{IJMPB} \textbf{19}, 1835 (2005).

\bibitem{VolovikPNAS} G. E. Volovik, \emph{PNAS} \textbf{97}, 2431 (2000).

\bibitem{WFmonopoles} Th. Busch, and J. R. Anglin, \emph{Phys. Rev. A}, \textbf{60}, 2669(R), (1999).

\bibitem{Antiferro} H. T. C. Stoof, E. Vliegen, and U. Al Khawaja, \emph{Phys. Rev. Lett}, \textbf{87}, 120407, (2001).

\bibitem{Atomic2009} V. Pietila, and M. Mottonen, \emph{Phys. Rev. Lett.} \textbf{103}, 030401 (2009).

\bibitem{Magnetricity2009} S. T. Bramwell et al., \emph{Nature} \textbf{461}, 956 (2009).

\bibitem{Microcavities} A. V. Kavokin, J. J. Baumberg, G. Malpuech, F. P. Laussy, \emph{Microcavities} (Oxford University Press, 2007).

\bibitem{Kasprzak2006} J. Kasprzak et al., \emph{Nature} \textbf{443}, 409 (2006).

\bibitem{AmoSuperfl} A. Amo et al, \emph{Nat. Phys.} \textbf{5}, 805 (2009).

\bibitem{DeveaudScience} K. G. Lagoudakis et al., \emph{Science} \textbf{326}, 974 (2009).

\bibitem{RecentVortexEvolution} K. G. Lagoudakis et al., \emph{Phys. Rev. Lett.} \textbf{106}, 115301 (2011).

\bibitem{Pigeon} S. Pigeon, I. Carusotto, and C. Ciuti, \emph{Phys. Rev. B} \textbf{83}, 144513 (2011).

\bibitem{ScienceAmoSoliton} A. Amo et al., \emph{Science} \textbf{3}, 1167 (2011).

\bibitem{ReviewSpin} I. A. Shelykh et al., \emph{Semic. Sci. Techn.} \textbf{25} 013001 (2010).

\bibitem{Renucci2005} P. Renucci et al., \emph{Phys. Rev. B} \textbf{72}, 075317 (2005).

\bibitem{FlayacHalfSoliton} H. Flayac, D.D. Solnyshkov, G. Malpuech, \emph{Phys. Rev. B} \textbf{83}, 193305 (2011).

\bibitem{RuboHV} Y. G. Rubo, \emph{Phys. Rev. Lett.} \textbf{99}, 106401 (2007).

\bibitem{Shelykh2006} I.A. Shelykh et al., \emph{Phys. Rev. Lett.} \textbf{97}, 066402 (2006).

\bibitem{APLMalpuech} G. Malpuech et al., \emph{Appl. Phys. Lett.} \textbf{88}, 111118 (2006).

\bibitem{Dasbach2005} G. Dasbach et al., \emph{Phys. Rev. B} \textbf{71}, 161308 (2005).

\bibitem{KavokinMeissner} Yu. G. Rubo, A.V. Kavokin, and I.A. Shelykh, \emph{Phys. Lett. A} \textbf{358}, 227 (2006).

\bibitem{RuboPolSplit} M. Toledo Solano, and Y. G. Rubo, \emph{J. Phys.}: Conf. Ser. \textbf{210} 012024 (2010).

\bibitem{movie} See EPAPS Document No. [number will be inserted later] for the HV propagation movie.

\bibitem{Leggett} A. J. Leggett, \emph{Quantum Liquids} (Oxford University Press, Oxford, 2008).

\bibitem{Leggett2} M. R. Matthews et al., \emph{Phys. Rev. Lett.} \textbf{83}, 3358 (1999).

%\bibitem{Weinfurtner} S. Weinfurtner et al, Phys. Rev. Lett. 106, 021302 (2011).

%\bibitem{Ciuti2004} I. Carusotto, C. Ciuti, Phys. Rev. Lett. 93, 166401 (2004).

%\bibitem{Wertz2010} E. Wertz et al, Nature Physics 6, 860 (2010).

%\bibitem{Tassone} F. Tassone and Y. Yamamoto, Phys. Rev. B \textbf{59}, 10830 (1999).

%\bibitem{Carusotto2008} R. Balbinot et al, Phys. Rev. A 78, 021603(R) (2008).

%\bibitem{CarusottoNJP} I. Carusotto et al, New Journ. Phys. 10, 103001 (2008).

%\bibitem{Wheeler} J.A. Wheeler, Geometrodynamics (Academic, New York, 1962).

%\bibitem{SuppVideo} See EPAPS Document No. [number will be inserted later] for a movie showing the evolution of the photon density.

%\bibitem{Lagoudakis2008} K. G. Lagoudakis et al., \emph{Nat. Physics} \textbf{4}, 706 (2008).

%\bibitem{Macher} J. Macher, and R. Parentani, \emph{Phys. Rev. A} \textbf{80}, 043601 (2009).

%\bibitem{Schutzhold} R. Schutzhold, W. G. Unruh, \emph{Phys. Rev. D} \textbf{78}, 041504(R) (2008).

%\bibitem{Mayoral2010} C. Mayoral et al., \emph{New J. Phys.} \textbf{13} 025007 (2011).

%\bibitem{VolovikHalfSoliton}  M. M. Salomaa, G. E. Volovik, \emph{J. Low Temp. Phys.} \textbf{4}, 319, (1988).

%\bibitem{VolovikReview} G. E. Volovik, \emph{The Universe in a Helium Droplet} (Clarendon Press - Oxford, 2003).

\end{thebibliography}
\end{document}